\documentclass{nature}

\usepackage{graphicx}% Include figure files
\usepackage{amsmath}
\title{Time-reversed wave mixing in nonlinear optics}

\author{Yuanlin Zheng$^{1}$, Huaijin Ren$^{1}$, Wenjie Wan,$^{1,2 *}$ and Xianfeng Chen$^{1 \dag}$}
\date{Dec. 13, 2012}

\begin{document}

\maketitle

\begin{affiliations}
 \item Department of Physics, Key Laboratory for Laser Plasmas (Ministry of Education),
Shanghai Jiao Tong University, Shanghai 200240, China
 \item University of Michigan-Shanghai Jiao Tong University Joint Institute, Shanghai Jiao Tong University, Shanghai 200240, China
\end{affiliations}

\begin{abstract}
Time-reversal symmetry is important to optics. In linear optics, a
time-reversed process to laser emission enables total absorption
of coherent light fields into an optical cavity of loss by
time-reversing the original gain medium. In nonlinear optics, time
symmetry exists for some well-known processes such as parametric
up/down conversion, sum/difference frequency generation, however,
combining exact time-reversal symmetry with nonlinear wave mixings
is yet explored till now. Here, we demonstrate time reversed wave
mixings for second harmonic generation (SHG) and optical
parametric amplification (OPA). This enables us to observe the
annihilation of coherent beams under time-reversal symmetry by
varying the relative phase of the incident fields. Our study
offers new avenues for flexible control in nonlinear optics and
potential applications in efficient wavelength conversion,
all-optical computing.
\end{abstract}

Time-reversal symmetry dominates in many physical systems; it
allows a physical process to reverse in a backward direction of
time. This powerful symmetry enables many practical applications
under such reversible principle, with examples including spin
reversing in nuclear magnetic resonance (NMR) imaging \cite{Hahn},
acoustic and electromagnetic focusing using time-reversal
mirror\cite{Kuperman, Fink, Fink2000, Lerosey}, phase conjugation
mirror in optics\cite{Yariv,Guang}, etc. Especially in optics,
light transmission in random medium can be enhanced by several
orders by exploring the time-reversal symmetry \cite{Mosk2010,
Mosk2011, Mosk2012}, similarly for enhancing second harmonic
signals in random medium\cite{Yaron2011}. Also a quantum splitter
can be obtained by time-reversed Hong-Ou-Mandel interferometer
\cite{Kumar}. Recently, a novel concept named ``coherent perfect
absorber" (CPA) \cite{Chong2010, Chong2, Wan} that explores
time-reversed process to laser emission has shown that: incident
coherent optical fields can be perfectly absorbed by a
time-reversed optical cavity by replacing the gain with equal
amount of loss. Also the incident fields and frequency should
coincide with those of the corresponding lasing modes with gain
under time symmetry. Since then, many efforts have been made to
study CPA properties with different geometries
\cite{Longhi,Longhi2010,Longhi2011,Longhi2012,Chong2010,Chong2011,Chong2,
Wan}. However, most of them remain in linear optics regime.
Nonlinear version of CPA has been theoretically proposed to
investigate signal and idler beams' phase varying dynamics in
presentence of pumping beam under time reversed optical parametric
oscillation (OPO) scheme \cite{Longhi}. These studies of time
symmetry have been attracting increasing attention, since it
provides alternative and substantial ways to manipulate light in
nonlinear regime. Here, we experimentally exam the time-reversal
symmetry for two of classical nonlinear wave mixing processes: SHG
and OPA, characterize their nonlinear properties as opposite to
their time-reversal counterparts, and reveal the nontrivial
dynamics of phase varying in time-reversed nonlinear wave mixing
schemes.

In nonlinear optics, several pairs of wave mixing phenomena seem
to be reversible in time. For example, sum-frequency generation
(SFG) converts two photons into one, natively the opposite
scenario, difference-frequency generation (DFG) that splits one to
two seems to be the counterpart to the former one in time.
However, time symmetry requires the exact inversion of all
physical parameters with respect to time. These pair processes
have dramatic different initial conditions in term of pumping
intensity, polarization, phase, etc. In order to achieve
time-reversal, one has to consider them all in time.

The SHG process converts a fundamental wave (FW) into its second
harmonic (SH) through nonlinear response of a medium. However, the
energy flow can oscillate between the fundamental and harmonic
waves, determined by the phase mismatch condition and phase
difference of the interacting waves. Under microscopic theme, this
energy oscillation between FW and SH results from interference
effects due to nonlinear polarizations of electrical dipoles. The
rates of such up and down conversions are determined by their
nonlinear susceptibilities, according to permutation symmetry
\cite{shen, Naguleswaran1998}, they are equivalent. The rest
conversion relies on the phase relationship. Below the depleting
pump limit, FW and SH can be described by two coupled linear
equations. Even reversed in time , this coupling still obeys the
same physical system. For a non-seeded scheme, SH starts to grow
from quantum fluctuations draining the energy from FW along the
propagation. The exact time reversal of the problem can lead SH to
none for the initial stage, similar to CPA in the linear case.

In this letter, we exam time reversed processes for SHG and OPA.
These backward nonlinear wave mixings allow the unique property of
annihilation of coherent beams in a nonlinear quadratic medium by
time reversal. Unlike the case of CPA in linear regime where
incident fields are totally absorbed and converted into heat
\cite{Chong2010, Longhi}, here annihilation of incident fields can
lead to the generation of new fields, such backward parametric
interactions may have a future for efficient wavelength conversion
for better long-wavelength detection, e.g. mid-IR, THz. More
interestingly, a flexible phase control can be achieved to probe
the nonlinear dynamics during the wave mixing, redirecting the
wave mixing in forward or backward time. This also offers new
avenues for flexible control in nonlinear optics and potential
applications in all-optical computing.

In an undepleted-pump SHG scheme, we consider the
quasimonochromatic waves with carrier frequencies of fundamental
wave at $\omega_1$, and second harmonic wave at
$\omega_2=2\omega_1$. After representing the electric field as
$\mathcal{E}_i(z,t)=E_i(z)\exp(-i\omega_i
t)+c.c=A_i(z)\exp[-i(\omega_i t-k_i z)]+c.c.$, where $i=1,2$
refers to FW and SH, respectively, and $A_i(z)$ is the slowly
varying amplitude (slow-varying envelope approximation), the wave
equations governing the spatial field envelope $E_i(z)$ in a
nonmagnetic nonlinear medium can be current charges expressed as
\cite{Boyd}:
\begin{equation}\label{Eq:coulpler}
    \frac{dA_2}{dz} = \frac{i\omega_2^2d_{eff}}{k_2 c^2}A_1^2e^{i\Delta
    kz},
\end{equation}
where the phase-mismatching vector $\Delta k=2k_1-k_2$. And
$d_{eff}$ is the effective nonlinearity, $c$ is the speed of
light. To consider its time-reversed counterpart, we take the
complex conjugate of Eq. \ref{Eq:coulpler} to get
\begin{equation}\label{Eq:coulpler_conj}
    \frac{dA_2^*}{dz} = e^{i\pi}\frac{i\omega_2^2d_{eff}}{k_2 c^2}{A_1^*}^2e^{i\Delta
    k'z}.
\end{equation}
Note that Eq. \ref{Eq:coulpler_conj} is the same as Eq.
\ref{Eq:coulpler} except for an extra phase difference
($\Delta\phi=\phi_2-2\phi_1=\pi$) and sign-flipped
phase-mismatching vector ($\Delta k'=-\Delta k$). The complex
conjugate fields $E_i^*(z)$, which represents the backwards
fields, satisfy Eq. \ref{Eq:coulpler_conj}. With such exact
time-reversal configurations, SHG beams can undergo a backward
process with respect to its counterpart in forward-time direction
under a time reversible environment that excluding magnetism and
loss. Here the model is based on monochromatic wave or waves with
simple envelopes, for more complex ones one has to time-reverse
theirs waveforms in space and time  as well, e.g. through
time-reversed mirror \cite{Fan} or nonlinear wave
mixing\cite{Miller}.

Generally, in the SHG process, SH signals grow along the
propagation in a nonlinear optical crystal; however, they can also
decay due to the familiar phase-matching problem. Here the phase
plays an important role: it directs the time axis where energy
flows. For example, for scheme SHG in quasi-phase-matching (QPM)
gratings, one can purposely flip the 2nd-order susceptibility
($\chi^{(2)}$) of ferroelectric domains adding a phase jump to SH
waves in order to prevent SH falling back to FW, encouraging SHG
conversion. Fundamentally, this originates from an interference
effect, similar to the linear CPA, however, such interference is
nonlinear one due to nonlinear polarization at molecular level
between SH and FW beams\cite{shen}, hence, the phase difference is
crucial here to consider the time-reversed process of a SHG as
well.
%For a spontaneously grown SHG (starting
%from zero), its time-reversed process points to exactly the
%opposite way: SH signals with the appropriate relative amplitude
%and phase to the pumping beam can be totally annihilated when they
%incident onto a nonlinear crystal.
%Fundamentally, this also origins from an
%interference effect, similar to the linear CPA, however, such
%interference is nonlinear one due to nonlinear polarization at
%molecular level between SH and FW beams.  This time reversal
%corresponds to interchanging incoming and outgoing fields, the
%time-reversed process of a SHG corresponds to perfect annihilation
%of incoming SH fields back to FW fields.

\begin{figure}[!htbp]
  \centering
  \includegraphics[width=0.7\textwidth]{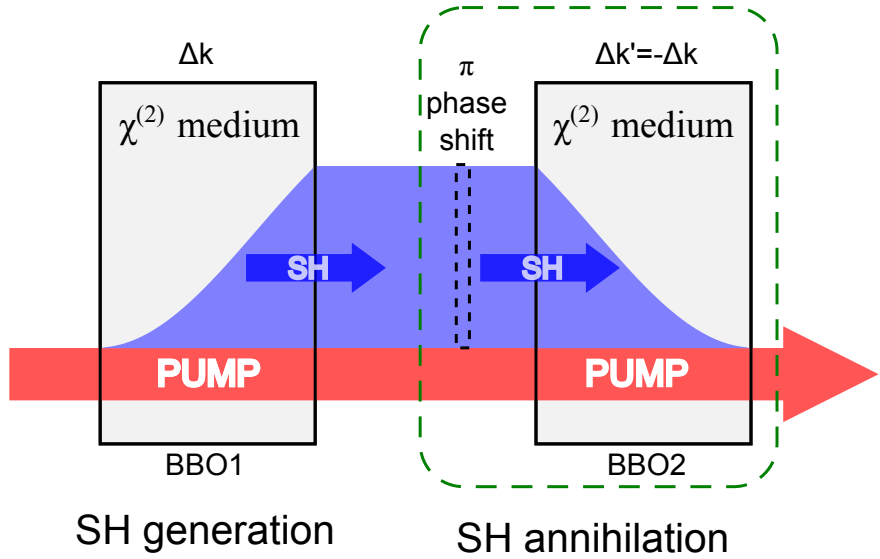}
  \caption{\small \textbf{The schematic of SH generation and annihilation.}
  The wave mixing behavior of SHG process in the first nonlinear medium is time-reversed in the second (outlined in dashed box).
  The condition for time-reversed symmetry is obtained by adding a $\pi$ phase shift to FW/SH phase difference and flipping the sign
  of phase-mismatching vector. For a spontaneously grown SHG (starting from zero) in BBO1, its time-reversed process leads to
  ``perfect" SH annihilation in BBO2 under time reversal
  symmetry.
  }\label{Fig:figure1}
\end{figure}

Figure \ref{Fig:figure1} shows the scheme for one time-reversed
SHG. Here a pair of two identical thin BBO (beta barium borate)
crystals cut for Type I SHG@1064 nm was used (see Supplement). The
crystal length $L$ is short enough to ensure pump non-depletion
(small signal approximation), i. e. $A_1(z)=A_1(0)$. The
corresponding phase-mismatching vectors $\Delta k$ ($\Delta k'$)
in BBO1 (BBO2) can be independently tuned by the rotating the
crystals. The intensity of the SH after BBO1 can be obtained by
direct integrating Eq. \ref{Eq:coulpler} to give \cite{Boyd}:
$I_2=2d_{eff}^2\omega_2^2I_1^2/(n_1n_2\epsilon_0c^2)L^2sinc^2(\Delta
kL/2)=I_{norm}sinc^2(\Delta kL/2)$, where $I$ is the wave
intensity, $n_i$ is refractive index and $\epsilon_0$ is the
vacuum permittivity. SH is generated when $\Delta kL/2$ is in the
range of $(-\pi, \pi)$, and the phase difference between the
generated SH and FW is $\Delta\phi=\Delta kL/2$. Thus BBO1 acts
dual-roles as an SH source generator and a phase controller. Both
the pump FW and generated SH are then directly incident into the
second nonlinear medium BBO2, where time-reversed SHG is
constructed. For a spontaneously grown SHG (starting from zero) in
BBO1, its time-reversed process points to exactly the opposite
way: SH signals with the appropriate relative amplitude and phase
to the pumping beam can be totally annihilated when they incident
onto a nonlinear crystal, as will be shown in BBO2.

\begin{figure}[!htbp]
  \centering
  \includegraphics[width=0.8\textwidth]{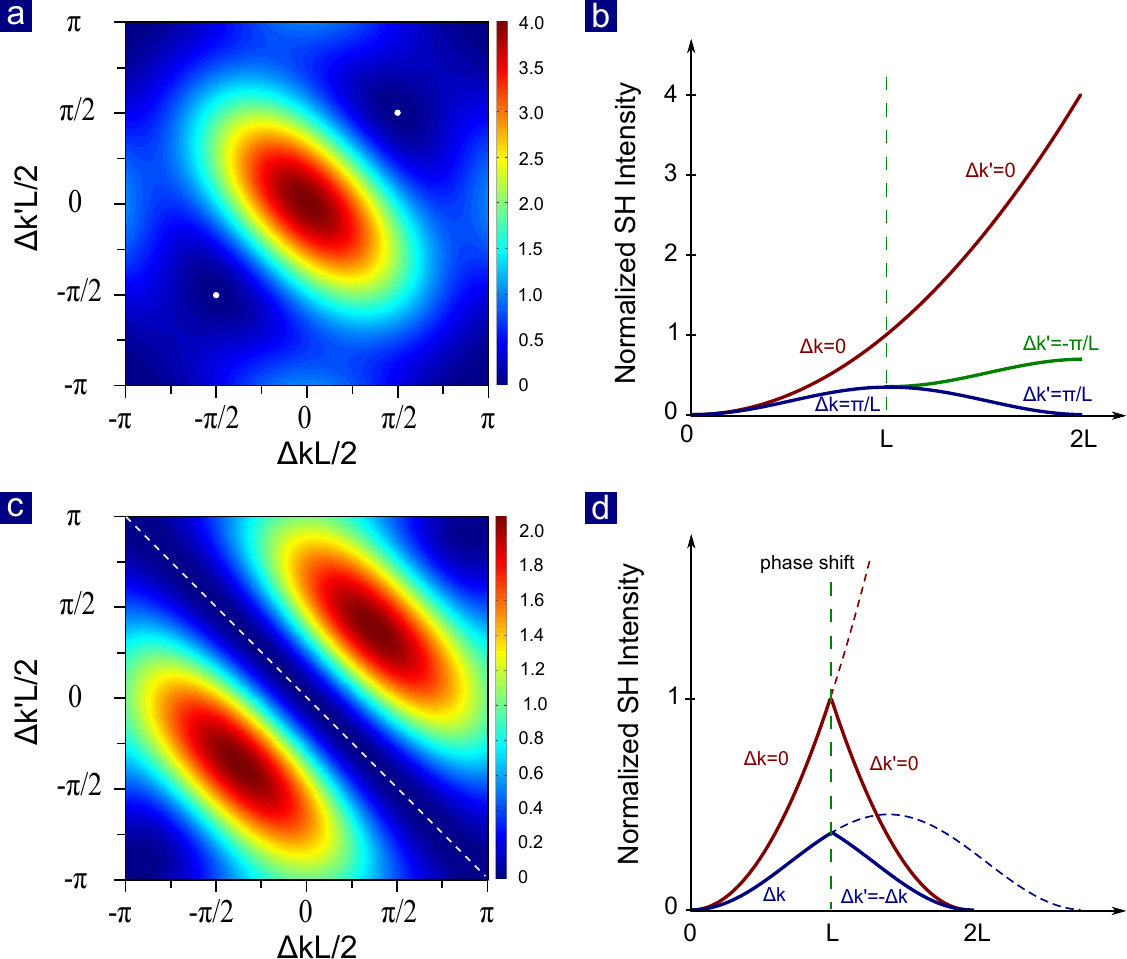}
  \caption{\small \textbf{The total SH output characteristics and the SH intensity along the propagation path.}
  \textbf{a}, The normalized total SH output without a phase shift between the two crystals. The two white dots show two situations when SH generated in BBO1
  is cancelled out in BBO2 due to dephasing. \textbf{b}, SH intensity along the propagation length in the two crystals
  in phase-matching and mismatching situations. \textbf{c}, The normalized total SH output is greatly modified after introducing an additional $\pi$ phase shift to $\Delta\phi$
  in between the two crystals. The diagonal dashed line shows total SH outputs equal to none, indicating any SH generated in the first nonlinear medium can always be
  cancelled out in the second. \textbf{d}, SH intensity along the propagation length in the two crystals. The SH intensity evolution manifests itself in spatially
  symmetric pattern, which is also an indication of time-reversal of the wave mixing process. White dash dots and line in \textbf{a} and \textbf{c} indicate zero intensity.
  }\label{Fig:figure2}
\end{figure}

To gain insight into the nonlinear waves mixing properties along
the proposed time-reversed structure, we calculate the total
output after BBO2 by integrating Eq. \ref{Eq:coulpler} along the
propagating path. We refer the normalized SH intensity $I_{norm}$
to unity hereafter, which equals to the output of BBO1 at
phase-matching condition. The total SH output is studied by
scanning through phase-matching vectors for both crystals, as
shown in Fig. \ref{Fig:figure2}a. The points of interest are those
where SH generated in BBO1 is totally cancelled out in BBO2
leaving only FW out of BBO2. In Fig. \ref{Fig:figure2}a, where the
two crystals are simply cascaded, SH vanishes if only $\Delta
kL=\Delta k'L=\pi/2$. However, this is well-known that SHG at
phase-mismatching condition induces energy flow oscillation
between FW and SH at the period of twice of the coherent length
$L_c=\pi/\Delta k$. The four corners are not the desired ones,
which are just due to non-SH input and also no SH generated by
BBO2. Figure \ref{Fig:figure2}b shows the typical SH intensity
along the propagating path at the conditions of phase-matched,
quasi-phase-matched, and phase-mismatched conditions (shown in
red, green and blue lines, respectively). The situation of green
line resembles the idea of QPM, which flips the sign of
phase-mismatching vector when FW and SH are out of phase at $L_c$
to gain a continuous SH generation. Note that, SH's intensity only
flows back under the phase mismatching condition, and occurs at
twice of the coherent length. Along its propagation, SH exhibits
symmetry between the two crystals. This stimulates the quest of
exploring time-symmetry for re-converting SH back to FW at any
moment despite their phase relations.
%[cite here]
%This is at the core of QPM mechanism
%in controlling wave mixing processes. The discussion here is
%well-known already.

According to previous analysis, time reversal symmetry can be
implemented by introducing an additional $\pi$ phase difference
before BBO2 and reversing the phase mismatching vector in BBO2. We
re-plot the SH output after BBO2 in Fig. \ref{Fig:figure2}c. As it
shows, the SH generated in BBO1 can always be cancelled out in
BBO2 whenever $\Delta k'=-\Delta k$ is fulfilled. This reveals the
circumstance where SHG generated by BBO1 undergoes annihilation
process along a time-reversal path in BBO2. Figure
\ref{Fig:figure2}d illustrates this idea better by showing SH
intensity along the propagating path across both crystals. With
these conditions, the intensity profiles of SH preserves the
spatial symmetric along the propagation axis through two crystals.
It is more convincing to observe such symmetry even under the
phase-matched condition in Fig. \ref{Fig:figure2}d. In all these
circumstances, the phase plays a core role in the determining of
the direction of the wave mixing process.

\begin{figure}[!htbp]
  \centering
  \includegraphics[width=1\textwidth]{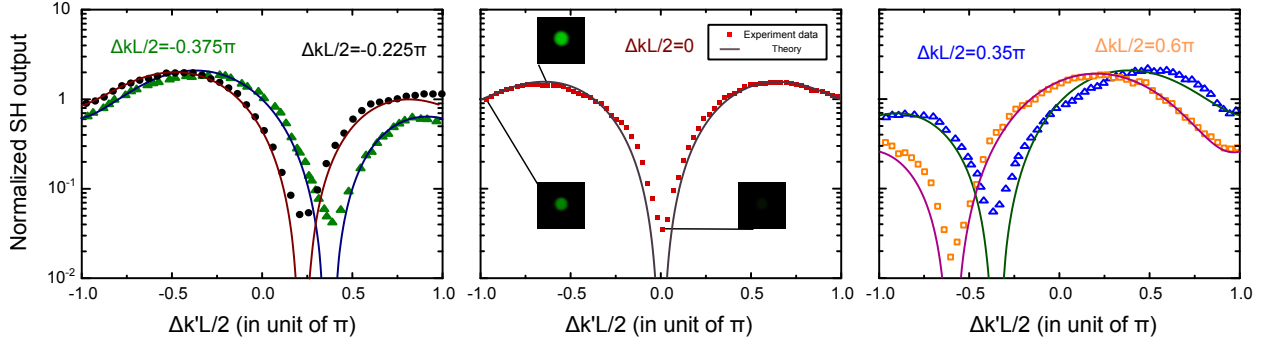}
  \caption{\small \textbf{Experimental measurements of reversed SHG.} The measured total SH intensity with different FW/SH inputs into BBO2
  by varying its phase-mismatching vector $\Delta k'$. The dips are the signature of CPA, where SH generated in BBO1 is significantly
  absorbed by time-reversed SHG in BBO2. (Insets) Photographs of the SH spot at the specified conditions.
  }\label{Fig:figure3}
\end{figure}

Experimentally, we exam this time-reversal SHG by considering
varying the phase mismatching vector to satisfy the condition for
total annihilation.  We set BBO1 for several different conditions
where $\Delta kL/2=-0.375\pi, -0.225\pi, 0, 0.35\pi, 0.6\pi$. The
SH intensity generated in BBO1 is $I_2^{in}=sinc^2(\Delta kL/2)$
with the FW/SH phase difference $\Delta\phi=\Delta kL/2$.  Then a
$\pi$ phase shift to $\Delta\phi$ is added between the two BBO
crystals (See Supplement). Figure \ref{Fig:figure3} shows the
measured SH output from BBO2 by scanning the phase mismatching
vector.  Significant annihilation of SH in BBO2 occurs when the
time-reversal requirements are met, that is, the dips of the SH
output curves are located at $\Delta k'=-\Delta k$. The results
clearly indicate that for an SHG even with low conversion
efficiency there always exists a symmetrically reversed SHG
process that is capable of ``perfect absorbing" its SH wave when
coherently illuminated by both FW and SH. The corresponding
attenuation to each dip of the curves is measured to be 11.5 dB,
12.4 dB, 14.5 dB, 11.4 dB and 10.6 dB, yet the SH generated in
BBO1 is predicted to be totally absorbed in BBO2
theoretically\cite{Longhi, Chong2010}. Such modulation depth is
limited by the experimental conditions, e.g. pumping intensity,
wavefront, coherence.
%In the time reversed SHG, the annihilation of SH can be seem as a
%procedure of total parametric down conversion. One photon of SH is
%``stimulated" to split to two, leading to total annihilation of
%SH. In the viewpoint of classical fields, it is possible to
%achieve perfect annihilation of SH wave by exact time reversal,
%since SH from SHG starts from none.
However, strictly complete annihilation cannot be reached should
quantum mechanism and quantum noise are
considered\cite{Longhi,Chong2012}.

\begin{figure}[!htbp]
  \centering
  \includegraphics[width=0.9\textwidth]{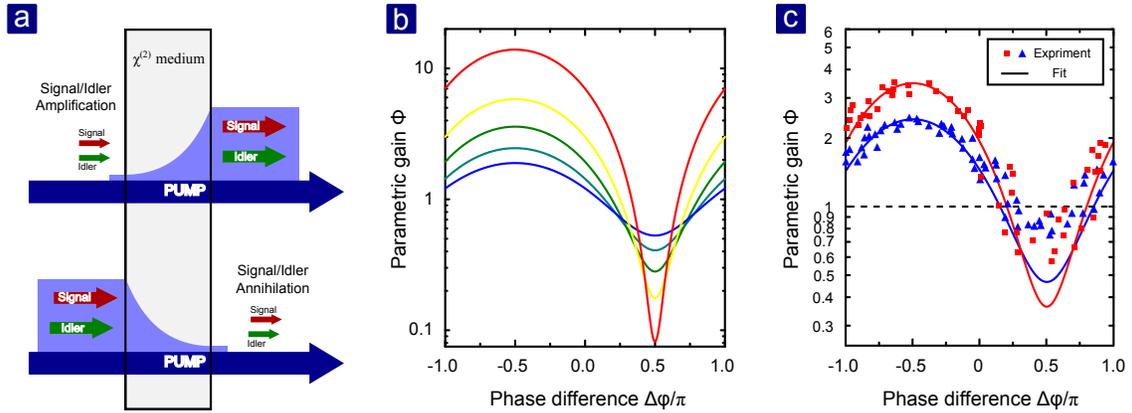}
  \caption{\small \textbf{The schematic of OPA and its reversal.} \textbf{a}, The schematic of Signal/Idler pair amplification and
  annihilation. \textbf{b}, Behavior of the parametric gain $\Phi$ vs. total phase difference $\Delta\phi$ for various pump
  intensities. \textbf{c}, The experimentally measured overall $\Phi$ vs. $\Delta\phi$. Gain variety along pump pulses is taken
  into account in data fitting.
  }\label{Fig:figure4}
\end{figure}

%%%%% ÒªÌí¼Ó¹ý¶É
This kind of time-reversal phenomena are rather general and should
be expected to occur for nonlinear optics with SHG in the
up-conversion region, but also another typical down-conversion
mixing -- OPA, which provides an important example in nonlinear
optics. Moreover, similar to two-channel CPA, the three waves
mixing in an OPA process also adds more complexity by considering
phases and gain. Compared with reversed SHG, the pumping effect
can plays a critical role in reversed OPA geometry. Here we extend
our study to a reversed two-channel OPA, which is qualitatively
requires two coherent input beams and the pumping beam.
Annihilation of waves can only be achieved when the relative
phase, amplitude and pumping level conditions are reached, as
schematically shown in Fig. \ref{Fig:figure4}a. Thus, it is not
only sensitive to frequency but also to the amplitude and phase of
the input light.

We consider non-depleted Type II degenerate OPA of three
quasimonochromatic waves with carrier frequencies of fundamental
wave at $\omega_p$, signal and idler at
$\omega_s=\omega_i=\omega_p/2$ in BBO crystal. By slow-varying
envelope approximation, neglecting group velocity mismatch (GVM)
and group velocity dispersion (GVD), the envelops satisfy the
coupled wave equations\cite{Boyd}
\begin{equation}\label{Eq:OPA}
\begin{aligned}
    \frac{dA_s}{dz} &=&
    \frac{2i\omega_s^2d_{eff}}{k_sc^2}A_pA_i^*e^{i\Delta kz},\\
    \frac{dA_i}{dz} &=&
    \frac{2i\omega_i^2d_{eff}}{k_ic^2}A_pA_s^*e^{i\Delta kz}.
%    \frac{dA_s}{dz} &=&
%    \frac{2i\omega_p^2d_{eff}}{k_pc^2}A_sA_ie^{-i\Delta kz}.
\end{aligned}
\end{equation}
The subscripts $p, s, i$ stand for pump, signal and idler,
respectively. The phase mismatching vector $\Delta k=k_p-k_s-k_i$
is required to be zero for efficient conversion. Hence, the
relative phases matters the most for now. $d_{eff}$ is the
effective nonlinearity and $c$ is the speed of light. One can find
the phase-matched parametric gain of signal/idler with respect to
pump intensity and total phase difference
($\Delta\phi=2\phi_p-\phi_s-\phi_i$) to be \cite{Boyd,Aariv_book,
Longhi}: $\Phi=\left|
cosh(gz)-i\exp(-i\Delta\phi)sinh(gz)\right|^2$, in which
$g=2\sqrt{(\omega_s\omega_i)/(n_sn_i)}d_{eff}A_p/c$ is
proportional to the pump amplitude. The results with different
pump intensities are shown in Fig. \ref{Fig:figure4}b. The gain or
attenuation is governed by both the pump intensity and the total
relative phase $\Delta\phi$. Hence, the relative phases matters
the most when we consider the time reversed OPA. Since $\Delta
k=0$, OPA and its time-reversal process is accessible in the same
condition, an OPA device can behave like an OPA-CPA
device\cite{Longhi}. Theoretically, the in-phase signal can be
amplified by the factor $\Phi$, whereas the out-of-phase signal is
attenuated by the same factor. The dip in the gain curve at
$\Delta\phi=\pi/2$ is the clear indicator of annihilation of
Signal/Idler beams when pumping above the threshold of OPA.

In our reversed-OPA experimental setup, the laser source is a
Nd:YAG nanosecond laser system which nominally delivers 4-ns
(FWHM) FW pulses at 1064 nm and 3-ns SH (see Supplement). The
polarization of attenuated FW is rotated to ensure balanced
orthogonally polarized Signal/Idler input onto the nonlinear
medium (BBO). The device is pumped by the SH and its phase is
finely tuned to adjust the total phase difference. In Fig.
\ref{Fig:figure4}c, we show the experimentally measured parametric
gain when the phase difference between the three waves is varied.
Far from $\Delta\phi=\pi/2$, the system behaves as an OPA
device($\Phi>1$). Note that, at $\Delta\phi=\pi/2$, a dip with
parametric gain $\Phi<1$ can be observed in the gain curves,
indicating the Signal/Idler annihilation, when pumping above the
threshold of OPA. Such a dip is the clear signature of ``colored"
OPA-CPA\cite{Longhi}, which is also the time-reversal process of
its counterpart OPA above threshold. In an ideal case, the product
of the gain factor for the in-phase and out-phase signal, i.e.
maximums and minimums on the curve, would be a unity. This is a
straightforward outcome of time-symmetry rule. However, there are
plenty of practical issues that would affect this relation. One
critical parameter is the pulse duration. The actual pulse of pump
wave with shorter time duration does not completely overlap with
the Signal/Idler pulses, meanwhile, the actual gain along the pump
pulse varies.
%which contribute to discrepancy between experimental and ideal theoretical model.
Nevertheless, these dips still clearly indicate that such
reversed-OPA device is capable of attenuating Signal/Idler given
the coherent illumination with correct phases, which behaves as a
``colored" OPA-CPA system. The result demonstrates two operating
states of a wave mixing process in its forward and time-reversed
direction.  However, in our experiment, gain depletion regime is
never reached. In this regime, OPA-CPA is still possible when
energy can be drained completely from the pumping beam
\cite{Longhi,Fejer2002}. This requires further studies in the
future.

In conclusion, we have demonstrated time reversed wave mixings for
SHG and OPA. This enables us to observe the annihilation of
coherent beams under time-reversal symmetry by varying the
relative phase of the incident fields. Time reversed SHG is able
to absorb coherent wave at second harmonic frequency of its pump.
For time reversed OPA, we show that the OPA can simultaneously
amplify and attenuate coherent signal and idler waves when pumped
above threshold. Our study provides a versatile platform for
flexible control in nonlinear optics and potential applications in
efficient wavelength conversion, all-optical computing.

%----------------------------------------------------------------%

%% Put the bibliography here, most people will use BiBTeX in
%% which case the environment below should be replaced with
%% the \bibliography{} command.
%\bibliography{CPA_reference}

%% Here is the endmatter stuff: Supplementary Info, etc.
%% Use \item's to separate, default label is "Acknowledgements"

\begin{addendum}
 \item This research was supported by the National Natural Science Foundation of China (Grant No. 61125503, 61235009),
 the National Basic Research Program 973 of China (Grant No. 2011CB808101), the Foundation for Development of Science
 and Technology of Shanghai (Grant No. 11XD1402600), the National 1000-plan Program (Youth), Shanghai Pujiang Talent Program (Grant
 No. 12PJ1404700)

 \item[Correspondence] Correspondence and requests for materials
should be addressed to Wenjie Wan (email: wenjie.wan@sjtu.edu.cn)
and Xianfeng Chen (email: xfchen@sjtu.edu.cn).
\end{addendum}

%%
%% TABLES
%%
%% If there are any tables, put them here.
%%

\end{document}